\documentclass[12pt]{article}
\usepackage{epsfig}

\def\beq{\begin{equation}}
\def\eeq#1{\label{#1}\end{equation}}
\def\eeqn{\end{equation}}

\def\beqa{\begin{eqnarray}}
\def\eeqa#1{\label{#1}\end{eqnarray}}
\def\eeqan{\end{eqnarray}}

\let\bar=\overbar

\def\Dslash{\not{\hbox{\kern-4pt $D$}}}
\def\dslash{\not{\hbox{\kern-2pt $\del$}}}

\def\msb{{\bar{\ssstyle M \kern -1pt S}}}

\newcommand{\gsim}{\raisebox{-0.7ex}{$\stackrel{\textstyle >}{\sim}$ }}
\newcommand{\lsim}{\raisebox{-0.7ex}{$\stackrel{\textstyle <}{\sim}$ }}

\def\Title#1{\begin{center} {\Large {\bf #1} } \end{center}}

\begin{document}

\Title{ Color Superconductivity and Blinking Proto-Neutron Stars}

\bigskip\bigskip

%+\addtocontents{toc}{{\it G. W. Carter}}
%+\label{CarterStart}

\begin{raggedright}  

{\it G. W. Carter\index{Carter, G.}\\
Department of Physics and Astronomy\\
State University of New York\\
Stony Brook, NY  11794-3800\\}
\bigskip\bigskip
\end{raggedright}

% Abstract for LANL: 
% If quark matter exists in the cores of neutron stars, it is 
% most likely color superconducting. Thus a phase transition
% from free to paired quark matter might occur during the first minute of 
% proto-neutron star evolution. In this talk I discuss how critical behavior 
% of the medium will modify neutrino diffusion, possibly leading to a short
% temporal variation in the neutrino signal detected on Earth. 

\section{Introduction}

Due to instabilities at the Fermi surface, matter composed of deconfined
quarks is believed to be a color superconductor at low temperatures
(see Refs.~\cite{Revs} and other reviews).
While the precise value of the energy gap at astrophysical 
densities is yet indeterminate, calculations using models of vacuum 
physics generally predict a gap in the neighborhood of 100 MeV for 
densities at and above three times that of equilibrium nuclear matter.
This implies that quark matter with a chemical potential fixed around 400
MeV undergoes a phase transition between a quark-gluon plasma state and one
of paired quarks at a critical temperature near 50 MeV.
In this talk I discuss how this could lead to a variation in the neutrino
signal emanating from a proto-neutron star formed in the wake of a supernova.

Our current understanding of Type II (core collapse) supernovae, 
based on a handful of observations and quite a bit of theoretical modeling,
begins with the implosion of the inner core of star of mass 8--20 
solar masses (see Ref.~\cite{PLPSR} for a recent review).
The evolution of the remnant proto-neutron star is driven by the diffusion
of neutrinos, which make their way through the hot ($T\sim 25$ MeV) and
dense ($n_B\sim 3 n_0$) core matter before a few are eventually detected
on Earth.
The temporal characteristics of this signal are determined for the most
part by the neutrino mean free path through this core.

For temperatures much lower than the superconducting energy gap, 
$T \ll \Delta$, quark states are replaced by diquark quasi-particles and
neutrino-quark cross sections are exponentially suppressed.
Thus at late times, when $T \sim 1$ KeV, the neutrino mean free path in quark
matter is practically infinite.
But within the first minute after core collapse, deleptonization heats 
the dense core to $T \simeq 50$ MeV before thermal neutrino emission cools
the system.
It is during this time that transitions between states of strongly-interacting
matter would occur.
While the phase transition from hadronic matter to a quark plasma is
likely first order (but poorly understood), a slightly later transition
from quark to superconducting quark matter might be second order, as in 
BCS theory.
The associated critical behavior of the latter would lead to a slowdown
in cooling, followed by a burst of streaming neutrinos as the temperature
eventually falls below the quark energy gap.
In what follows I will discuss inelastic neutrino-quark scattering in a
simplified proto-neutron star environment in order to quantify this 
possible phenomena and speculate on an observable signal.

\begin{figure}[bt]
\centering
{\epsfig
{figure=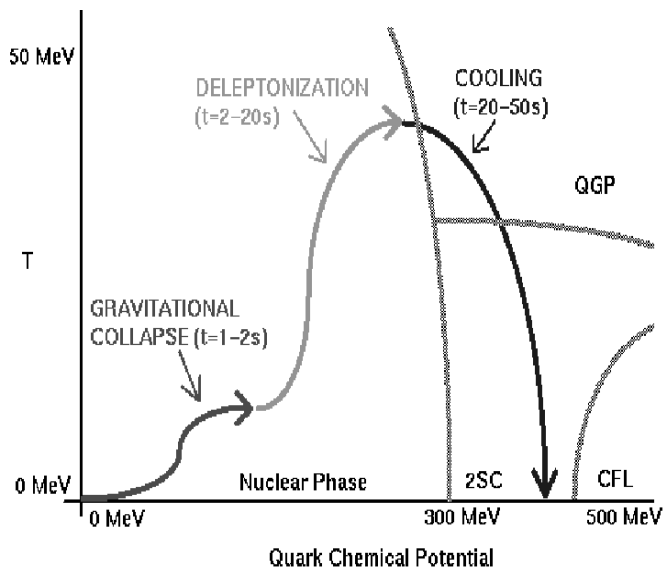,width=0.6\textwidth,height=0.49\textwidth}}
\caption{ A tour of the phase diagram via a proto-neutron star.  }
\label{pdiag}
\end{figure}

We are concerned here with the period in the proto-neutron star's evolution
during which the matter is at its hottest and most dense, immediately following
the Joule heating due to primordial lepton release.
The ambient conditions are a temperature near 50 MeV and a density
around four times that of nuclear matter, or equivalently a quark chemical
potential of 400 MeV.
This is the final period (Cooling) in the evolution mapped on the generic 
phase diagram for nuclear matter in Figure~\ref{pdiag}.

\section{Neutrino Mean Free Path in a Color Superconductor}

Thermal neutrino diffusion is the primary means of heat escape from the
dense core.
While noting that the neutrino production rate will also differ from 
that of normal matter \cite{JP}, in the diffusive
regime the dominant critical behavior will be a change in the
{\it inelastic} quark-neutrino cross section, since here neutrino production
rates decouple from the transport equation and depend only on the neutrino mean
free path.  
We thus calculated the differential and total cross sections in order to 
compute the neutrino mean free path in two-flavor quark matter.
The magnitude of the superconducting gap, $\Delta$, was taken as
arbitrary within a range of values found in recent literature.  
Closely following BCS theory, we assumed the gap to be a constant, and 
calculated the response functions and neutrino cross sections in the weak 
coupling approximation.

Details, including polarization operators, 
differential and total cross sections, and limiting behavior can be found in
Ref.~\cite{CR}.
The essential physical difference between neutrinos scattering in
superconducting quark matter as opposed to free quark matter is that, in
the quasi-particle phase, the incident neutrino must transfer energy greater
than twice the gap, or $E_\nu \geq 2 \Delta$, to excite a quark pair.
Stated equivalently,
when the gap becomes large ($\Delta \gsim T \sim 30$ MeV) the spacelike
response functions ($q_0 < \vert\vec{q}\vert$) are substantially suppressed,
and the inelastic cross section is greatly diminished.

\begin{figure}[t]
\centering
\leavevmode
{\epsfig
{figure=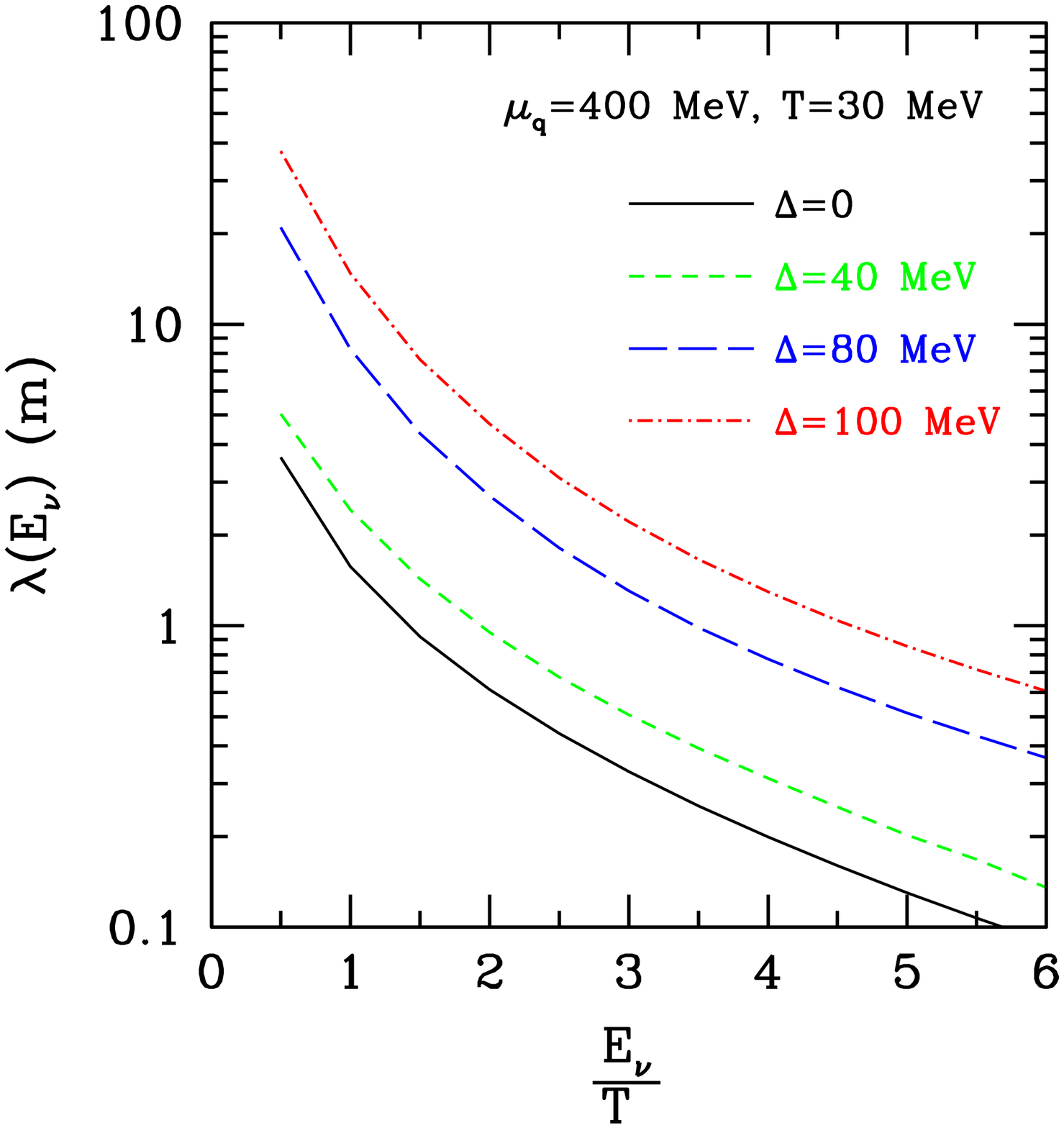,width=0.47\textwidth,height=0.49\textwidth}}
\leavevmode
{\epsfig
{figure=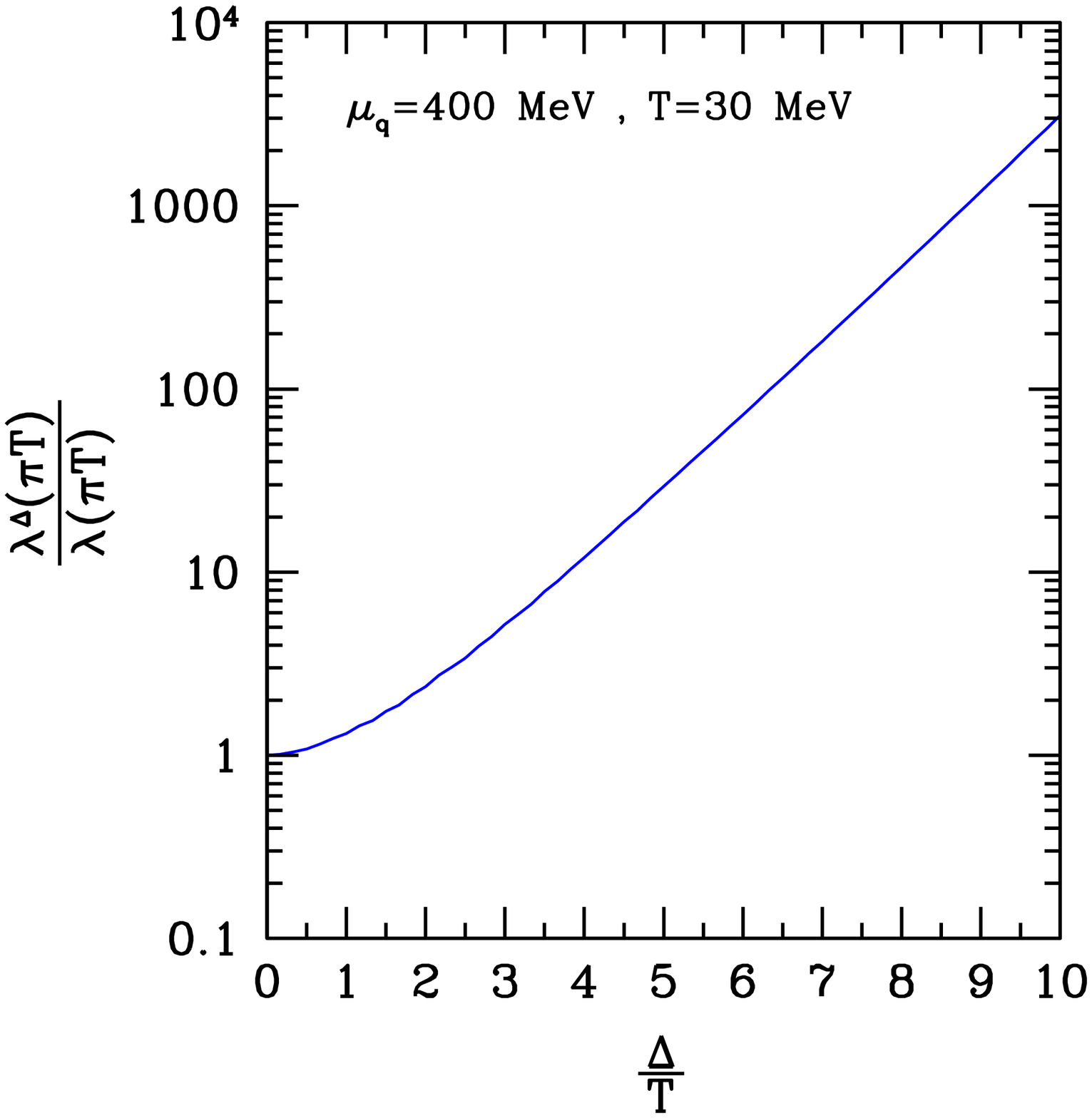,width=0.47\textwidth,height=0.49\textwidth}}
\caption[]
{Left Panel: Neutrino mean free path as a function of neutrino energy.
Right Panel: The neutrino mean free path for neutrino energy $E_\nu=\pi T$,
plotted as a function of the gap $\Delta$.
}
\label{lambda_fig}
\end{figure}

The mean free path, $\lambda = (\sigma/V)^{-1}$,
therefore grows with energy gap.
This main result is shown in Figure~\ref{lambda_fig}.
The left panel assumes a constant energy gap $\Delta$ at a constant temperature
$T=30$ MeV.
As expected, the mean free path grows substantially when $T < \Delta$ 
for all neutrino energies.  
In the right panel an average thermal neutrino energy of $E_\nu = \pi T$ 
has been assumed, and the superconducting gap is varied for fixed $T$.
The expected exponential behavior is clear for $\Delta \gsim 3 T$ and, combining
information from the two plots, one can deduce that for low temperatures the
neutrinos are able to pass through kilometers of superconducting quark matter 
with only an occasional collision.
These results are essentially independent of temperature for all
$T \lsim 50$ MeV.

\section{Cooling of an Idealized Quark Star}

The theoretical state of the art in modeling proto-neutron star cooling 
involves complex computer simulations of multi-phase environments.
To explore the consequences of the onset of superconductivity in a 
macroscopic core of deconfined quark matter, we will simply assume the
existence of such a system and approximate it by a sphere of quark matter.
We furthermore consider the relatively simple case of two massless flavors
with identical chemical potentials and disregard for now any 
non-superconducting quarks which might be present in the system.
We also make the (safer) assumption that the neutrino mean free path
is much smaller than the dimensions of the astrophysical object and several
orders of magnitude greater than the quark mean free path, meaning that
the system cools by neutrino diffusion.

In a such a sphere of quark matter, the diffusion equation for energy transport
is
\begin{eqnarray}
C_V\, \frac{dT}{dt}=-\frac{1}{r^2} \frac{\partial L_{\nu}}{\partial r} \,,
\label{ediff}
\end{eqnarray}
where $C_V$ is the specific heat per unit volume of quark matter,
$T$ is the temperature, and $r$ is the radius.
The neutrino energy luminosity for each neutrino type, $L_\nu$,
depends on the neutrino mean free path and the spatial
gradients in temperature.  
It is approximated by
\begin{equation}
L_\nu \cong -6\int dE_\nu\,\frac{c}{6\pi^2}\, E_\nu^3 r^2 \lambda(E_\nu)
\frac{\partial f(E_\nu)}{\partial r} \,,
\label{eflux}
\end{equation}
where $c$ denotes the speed of light in vacuum.
In our analysis we assume that neutrino interactions are dominated by the
neutral current scattering common to all neutrino types.
Consequently, we take the same neutrino and anti-neutrino mean free path
for every neutrino flavor, giving rise to the factor of six
in Eq.~(\ref{eflux}). 
The equilibrium Fermi distribution, $f(E_\nu)$, and the (scattering) mean
free path, $\lambda(E_\nu)$, are integrated over all neutrino energies,
$E_\nu$.

The solution to the diffusion equation will also depend on the size of the
system, the radius $R \sim 10$ km.
The temporal behavior is characterized by a time scale $\tau_c$,
which is proportional to the inverse cooling rate and can hence be deduced
from Eq.~(\ref{ediff}).
The characteristic time
\begin{equation}
\tau_c(T)= C_V(T) \frac{R^2}{c\langle\lambda(T)\rangle} \,,
\label{tauc}
\end{equation}
is a strong function of the ambient matter temperature since it
depends on the matter's specific heat and the neutrino mean free path.  This
applies to a system characterized by the radial length $R$ and the
energy-weighted average of the mean free path, $\langle\lambda(T)\rangle$.
Following our general treatment of the superconducting gap, we assume that the
temperature dependence of the specific heat is described by BCS theory.  We
will then use the scattering results to calculate
$\langle \lambda(T)\rangle$.  Furthermore, since neutrinos are in thermal
equilibrium for the temperatures of interest, we may assume
$\langle\lambda(T)\rangle \simeq \lambda(E_\nu=\pi T)\,$. 
Finally, we note that the diffusion approximation is only valid when $\lambda
\ll R$ and will thus fail for very low temperatures, when $\lambda \sim R$.

\begin{figure}[bt]
\centering
{\epsfig
{figure=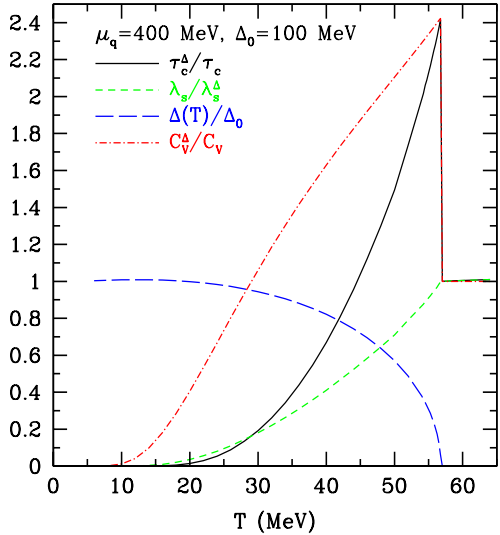,width=0.47\textwidth,height=0.49\textwidth}}
\caption[]
{
The extent to which different physical quantities are affected
due to the superconducting transition.  Ratios of the cooling time scale (solid
curve), the inverse mean free path (short-dashed curve) and the matter specific
heat (dot-dashed curve) in the superconducting phase to that in the normal
phase is shown as a function of the matter temperature. The ratio of the gap to
it zero temperature value $\Delta_0$ is also shown (long-dashed curve). The
quark chemical potential is $\mu_q=400$ MeV and $\Delta_0=100$ MeV. 
}
\label{tauc_fig}
\end{figure}

\section{A Signal of Color Superconductivity?}

The results of our calculations are summarized in Figure~\ref{tauc_fig}.
We have taken the temperature dependence of the gap (long dashed curve)
from BCS theory,
$\Delta(T)/\Delta_0 = \sqrt{1-(T/T_c)^2}$, where $\Delta_0$ is the
zero-temperature gap at a given density.
The specific heat (dot-dashed curve) is also taken from the mean-field result, 
and is peaked around $T=T_c$ as expected in a second-order phase transition.
With our result for the neutrino mean free path (short dashed curve), we 
combine factors to obtain the characteristic diffusion time (solid
line), shown as a ratio to that found in a system of free quarks.

The consequences of these results are interpreted as follows.
The early cooling rate,
around $T_c$, is influenced mainly by the peak in the specific heat associated
with the second order phase transition (the neutrino mean free path is
not strongly affected when $\Delta \ll T$).  
Subsequently, as the matter cools, both $C_V$ and $\lambda^{-1}$ decrease 
in a non-linear fashion for $\Delta \sim T$.  
Upon further cooling, when $\Delta \gg T$, both $C_V$ and $\lambda^{-1}$
decrease exponentially.  
Both of these effects accelerate the cooling process at later times.

We conclude that if the core of a neutron star formed in the aftermath
of a supernova contains a large amount of quark matter at sufficiently high
temperature and density, and if the associated neutrino signal can be 
detected on Earth, a unique temporal profile would be observed.
Specifically, we expect
a brief interval during which the cooling would
slow around $T\sim T_c$, signified by a period of reduced neutrino detection.
This would be followed by a brief burst of neutrinos after the temperature
falls well below $T_c$ and the quasi-quarks decouple from the thermal
neutrinos.

We note that many real-world complications would weaken or disrupt this signal,
as explained in Ref.~\cite{CR}.
Primary among them are the presence of strange quarks and the likely existence 
of a hadronic shell of nuclear matter covering the core, opaque to neutrinos, 
which would even out a sharp signal.
Furthermore, ours is in fact a generic prediction for any second-order 
phase transition, applied here to color superconductivity given our
understanding of proto-neutron star evolution and the phase diagram of
QCD.
However, given the real prospect of detecting neutrinos emitted from a 
future supernova event, transport processes like the ones we discuss here 
might someday serve as a reliable probe of the properties of extremely dense
matter.

\bigskip
I am pleased to thank Sanjay Reddy, my collaborator in the work described 
here.
I am also grateful to the organizers of this conference and the staff of 
Nordita for their hospitality and support.
This work was also supported by USDOE Grant DE-FG02-88ER40388.

\end{document}